# Title: Discovery of mesoscopic nematicity wave in iron-based superconductors

**Authors:** T. Shimojima[1,†,*], Y. Motoyui[2,†], T. Taniuchi[2,3], C. Bareille[2,3], S. Onari[4], H. Kontani[4], M. Nakajima[5], S. Kasahara[6,+], T. Shibauchi[7], Y. Matsuda[6], S. Shin[2,3,8,*]

**Affiliations:** [1]RIKEN Center for Emergent Matter Science (CEMS), Wako 351-0198, Japan

[2]Institute for Solid State Physics (ISSP), The University of Tokyo, Kashiwa, 277-8581, Japan

[3]Material Innovation Research Center (MIRC), The University of Tokyo, Kashiwa, Chiba 277-8561, Japan.

[4]Department of Physics, Nagoya University, Furo-cho, Nagoya 464-8602, Japan

[5]Department of Physics, Osaka University, Toyonaka, Osaka, 560-0043, Japan

[6]Department of Physics, Kyoto University, Kyoto 606-8502, Japan

[7]Department of Advanced Materials Science, The University of Tokyo, Kashiwa, 277-8561, Japan

[8]Office of University Professor, The University of Tokyo, Kashiwa, Chiba 277-8581, Japan

*Correspondences

†Equal contributions

+Present address: Research Institute for Interdisciplinary Science, Okayama University, 3-1-1 Tsushimanaka, Kita-ku, Okayama 700-8530, Japan.

**Abstract:**

Nematicity is ubiquitous in the electronic phases of iron-based superconductors. The order parameter that characterizes the nematic phase has been investigated in momentum space, but its real-space arrangement remains largely unexplored. We use linear dichroism (LD) in a low-temperature laser-photoemission electron microscope to map out the nematic order parameter of nonmagentic FeSe and antiferromagnetic $BaFe_2(As_{0.87}P_{0.13})_2$. In contrast to structural domains, which have atomic-scale domain walls, the LD patterns in both materials show peculiar sinusoidal waves of electronic nematicity with wavelengths more than 1000 times longer than the unit cell. Our findings put strong constraints on the theoretical investigation of electronic nematicity.

**Main Text:**

Electronic nematicity(*1*), a spontaneous breaking of the four-fold ($C_4$) rotational symmetry, has been observed in high-transition-temperature ($T_c$) superconductors, such as cuprates(*2,3*) and iron-based superconductors (IBSCs)(*4,5*). The nematicity has been suggested to be closely linked to the pseudogap state in cuprates(*6*) and the formation of Cooper pairs in IBSCs(*7*). In IBSCs, the electronic instabilities caused by spin and orbital degrees of freedom can induce lattice distortion from tetragonal to orthorhombic structure at $T_s$, referred to as nematic order(*8*). FeSe and $BaFe_2(As_{1-x}P_x)_2$ (AsP122) are representative of IBSCs exhibiting nematic order. The former shows nematicity without any long-range magnetic order(*9*) (Fig. 1A), whereas the latter exhibits a nematic order slightly above the antiferromagnetic (AFM) order ($T_N$)(*10*) (Fig. 1B).

Electronic nematicity in IBSCs is associated with the anisotropy of the in-plane band structure near the Fermi level ($E_F$), composed of multiple Fe 3*d* orbitals(*11*). As depicted in Fig. 1C1, the hole bands around Γ have *xz* and *yz* orbital components, which are energetically degenerate in the tetragonal state (*12*). In the nematic state, the energy level of *xz* band ($E_{xz}$) is ~10 meV higher than that of *yz* orbital ($E_{yz}$) at $\Gamma$(*13-15*), giving rise to the orbital polarization in the electronic density of states (DOSs) (Fig. 1C2). Anisotropy in inplane physical properties has also been reported from the transport(*4*) and nematic susceptibility measurements(*7,16*), suggesting that the nematicity is electronic in origin(*8,17*). In real space, optical microscopy measurements(*18-20*) reported nematic domains of 10 - 100 µm that coincided with the locations of local strain(*19*) and orthorhombic domains(*18,20*). Although these studies provided an overall picture of electronic nematicity, its spatial variation on a nanometer scale has been unclear.

Here we report a real-space visualization of the nematic order parameter on the crystal surface of FeSe and under-doped AsP122 by employing linear dichroism (LD) mapping in high-resolution laser-photoemission electron microscope (PEEM) (*21*) down to lowest temperatures.

By employing LD in laser-PEEM, we map the amplitude of the orbital polarization between *xz* and *yz* around $\Gamma$, owing to the selection rules (*12*). The LD signal is detected by taking the difference between the PEEM images obtained by laser beams whose polarization directions are parallel and perpendicular to the orthorhombic axes. As shown in Fig. 1, D and E, the LD images clearly reveal complex stripe patterns in the paramagnetic (PM)-nematic state of FeSe and the AFM-nematic state of $BaFe_2(As_{0.87}P_{0.13})_2$, covering the entire field of view. By inspecting the LD signals integrated in the selected LD images in Fig. 1, F and G, we find characteristic spatial variations. Near the boundaries at which the stripe patterns are rotated by 90 degrees (thick dashed lines in Fig. 1, D and E), plateaus appear at the maximum and minimum of the LD signal (left region of Fig. 1F). Away from the boundaries, the period of the signal becomes shorter and the plateaus gradually disappear, leading to a sinusoidal-wave-like profile (right region of Fig. 1F). The amplitude of the LD signal is reduced in the region where the wavelength is shorter, as shown in Fig. 1G and Fig. S1 (*12*). These peculiar features are commonly observed for BaFe$_2$(As$_{0.87}$P$_{0.13}$)$_2$ and FeSe (Fig. S2) (*12*).

To understand these features, we consider two scenarios, electron density wave and nematic domain walls. The former naturally explains the sinusoidal waveform in the LD signal. However, within this picture it is difficult to understand the plateaus and the period ranging from 400 nm to 1300 nm (Fig. S1) (*12*). In contrast, for the latter scenario, the complex stripe patterns (Fig. 1, D and E) are naturally understood because of a resemblance to the orthorhombic twinned domains(*18*). In the following, we show that this phenomenological model consistently explains the spatial variations in the waveform, period and amplitude of the LD signals.

In the following we assume a nematic domain wall much thicker than the typical orthorhombic twin boundary (TB) (5 nm in ref.*22*). According to the Ginzburg-Landau theory (*12*), the profile of the nematic domain wall can be expressed as $W(x)=D\tanh[x/2\xi_{nem}]$. The nematic domain wall has an intrinsic height $D$ and coherence length $\xi_{nem}$, both of which are assumed to be the same on the entire crystal surface as indicated by the black thick curves in Fig. 2, A to C. In this picture, the plateaus in the LD signal show up if the nematic-domain-wall interval [$S(x)$] is larger than $\sim 3\xi_{nem}$ (Fig. 2A, case I). As $S(x)$ decreases ($\sim \xi_{nem}$), neighboring domain walls are partially merged and then become nearly equivalent to a sinusoidal wave (Fig. 2B, case II). When $S(x)$ becomes comparable to $\xi_{nem}/2$, the LD amplitude starts to decrease while keeping the sinusoidal waveform (Fig. 2C, case III).

The LD signals in the gray areas of Fig. 1, F and G, are well fitted by the nematic domain wall model. By setting common parameters of $D = 0.13$ and $\xi_{nem} = 450$ nm, the fitting function composed of the train of nematic domain walls with $S(x)$ ranging from 400 to 1400 nm (green curve in Fig. 2D) reproduced the plateaus and sinusoidal waveform in the LD signal (Fig. S3) (*12*), corresponding to case I and case II, respectively. The low-amplitude LD signal in Fig. 2E is also reproduced by the nematic domain wall model with $D = 0.13$, $\xi_{nem} = 450$ nm and $S = 260$ nm (corresponding to case III). We note that the green curve composed of the nematic domain walls in Fig. 2E is equivalent to a sinusoidal wave with the wavelength of $2S(x)$. For FeSe, the LD signals can also be well fitted by the nematic domain wall model with coherence length $\xi_{nem} = 550$ nm (Fig. S2) (*12*). These results suggest that the observed peculiar sinusoidal LD signals (nematicity wave) can be attributed to nematic domain walls with a material-dependent mesoscopic coherence length.

Although the shape and the period of the LD signals in FeSe and BaFe$_2$(As$_{0.87}$P$_{0.13}$)$_2$ bear a close resemblance to each other, they exhibit notable differences in their temperature (*T*) dependence. In FeSe, which has no AFM order, the LD signals disappear at $T_s$ (90 K) and no discernible periodic signal is observed above $T_s$ as shown by the LD images (Fig. 3A) and the integrated LD signals (Fig. 3B). As depicted in Fig. 3C, *D* obtained from the fitting analysis (Fig. S4 and S5) (*12*) rapidly increases below $T_s$ as expected from the evolution of the *xz/yz* orbital polarization at Γ detected by bulk-sensitive laser-ARPES (Fig. S6) (*12*), which is consistent with the second-order tetragonal-to-orthorhombic phase transition(*9*). In stark contrast to FeSe, the amplitude of the LD signal for BaFe$_2$(As$_{0.87}$P$_{0.13}$)$_2$ remains finite even above $T_{s,N} = 93$ K. The most notable feature is that the sign of the LD signal is reversed at $T_{s,N}$, although the wavelength, and its nodal positions are unchanged with *T* as indicated by the dotted lines in Fig. 3D and crossing points of the LD signals in Fig. 3E. Figure 3F indicates the *T* evolution of *D* showing a characteristic *T* where the LD signals start to appear below $T^* \approx 120$ K. Previous experimental studies showed that the 122 systems exhibit the signatures of the electronic nematicity even above $T_{s,N}$. The LD signal at $T_{s,N} < T < T^*$ can be attributed to the lifting of the degeneracy of the *xz/yz* orbital bands as probed by ARPES on BaFe$_2$As$_2$ (*15*) and AsP122 system(*24*). In torque magnetometry data on BaFe$_2$(As$_{0.86}$P$_{0.14}$)$_2$, the $C_2$ component appears at temperatures similar to $T^*$ (*5*). The sign change across $T_{s,N}$ may be caused by ($\pi$,0) band folding due to the AFM order, which induces large inequivalency in the partial density of states between *xz* and *yz* electrons(*24*). These results suggest that the LD signals in both materials reflect the electronic nematicity in bulk crystals, regardless of the presence or absence of the AFM order.

A salient feature is that the LD signal of BaFe$_2$(As$_{0.87}$P$_{0.13}$)$_2$ shows up at $T_{s,N} < T < T^*$. For an ideal tetragonal lattice, an oscillating LD signal can be interpreted as the alternation of *xz*-dominated ($E_{xz} - E_{yz} > 0$) and *yz*-dominated ($E_{xz} - E_{yz} < 0$) electronic states (Fig. 4A). In real materials, short-range orthorhombic distortion detected above $T_s$(*5,25-27*) might be coupled to the nematicity wave. As *T* decreases below $T_s$, orthorhombic TBs appear to relax the lattice distortion. If the TB does not locate at each node of the nematicity wave, a sinusoidal LD signal can be interpreted by the sinusoidal modulation between *xz*-dominated ($E_{xz} - E_{yz} > 0$) and *yz*-dominated ($E_{xz} - E_{yz} < 0$) electronic states within a single orthorhombic domain (Fig. 4B). This situation is compatible with the characteristic dimension of the orthorhombic domain pattern on the order of 10 μm reported from optical measurements(*18-20*). Within this picture, the *xz/yz* orbital polarization can change sign without the change in the sign of orthorhombicity, challenging our understanding of the interplay between the orbital and elastic degrees of freedom (*8,17*). Alternatively, we consider the scenario in which each TB corresponds to a nematic domain wall (Fig. 4C). In this case, the orbital polarization should have the same sign in the orthorhombic setting of each domain (blue orbital in Fig. 4C), given that the orthorhombic axes are rotated by 90 degrees across the TB. We emphasize that regardless of interpretation, the observation of the mesoscopic nematicity wave suggests a nematic order parameter that exhibits high stiffness against real-space modulation and unusual decoupling from the lattice. In order to arrive at the correct real-space picture, experimental investigation of the orthorhombic domains with nanometer resolution will be highly important.

The interval of the nematic domain walls is determined by the repulsive force between each wall, which is expected to depend on the local environment. The origin of the unusually long coherence length remains unclear. The $T$-dependent LD signals for FeSe and BaFe$_2$(As$_{0.87}$P$_{0.13}$)$_2$ can be reproduced by assuming a constant value of $\xi_{nem}$ for each material as shown in Fig. S4 and S5 (*12*). These results suggest that the coherence length is not very sensitive to the band structure reconstruction at $T_s$ and $T_{s,N}$. It is also important to discuss the influence of local strain (*12*). In the $T$-cycle LD-PEEM measurements in Fig. S8 (*12*), once the nematic wave completely disappears by increasing $T$, it never appears at the same location at low $T$. This is to be contrasted with the strong pinning of the nematic domain boundary (50 - 100 μm) to the local strain (*19*). There is a possibility that the present LD-PEEM is detecting the central region of the micrometer-size nematic domains away from the local strain. Indeed, the slow development of the nematic order parameter below $T^*$ in LD-PEEM (Fig. 3F) is quite similar to the $T$ dependence of the photo-modulation response far away from the nematic domain boundary in the optical measurements (*19*).

Although the possibility of coexistence between mesoscopic and atomic-scale electronic nematicity remains (*27*), our results open the door to experimental and microscopic theoretical investigations to address whether the nematicity wave and its fluctuations contribute to the electron pairing in the IBSCs. In a wider context, mesoscopic phenomena have been found in other strongly correlated electron systems (*28-31*). The mesoscopic electronic modulations accompanied with hysteresis in the resistance (*32*) and colossal magnetoresistance effect (*33*) have been reported for several manganese oxides. The nematicity wave we observed may also give rise to interesting transport properties (*20*).

**Acknowledgments: Funding:** This work was supported by Grants-in-Aid for Scientific Research (Nos. JP18H01175, JP21H04443, JP18H05227, JP19H00649, JP19H00651) and on Innovative Areas "Quantum Liquid Crystals" (Nos. JP19H05824, JP19H05825) from Japan Society for the Promotion of Science and by JST CREST (JPMJCR19T5).; **Author contributions:** S.S. conceived the project. Y.Mo, T.T. performed the laser-PEEM measurements. T.Shim, Y.Mo analyzed the data. M.N. synthesized BaFe$_2$(As$_{0.87}$P$_{0.13}$)$_2$ single crystals. S.K., T.Shibauchi and Y.Ma synthesized FeSe single crystals. S.O. and H.K. provided the calculations for the band structure and PEEM signal. T.Shim wrote the paper with inputs from all coauthors. ; **Competing interests:** Authors declare no competing interests; and **Data and materials availability:** Data for this report are available at (*35*).

**Supplementary Materials:**

Materials and Methods

Supplementary Text

Figs. S1 to S8

References

**Figure captions:**

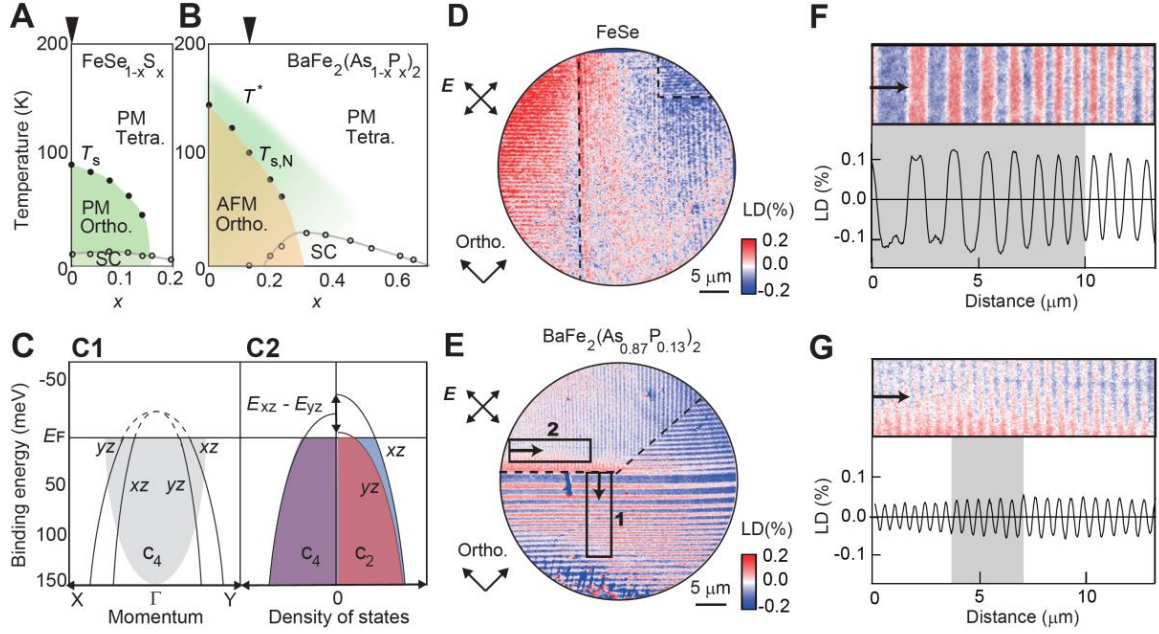

**Fig. 1. Laser-based LD-PEEM in the nematic ordered state.** (**A** and **B**) Phase diagrams of FeSe$_{1-x}$S$_x$(*34*) and BaFe$_2$(As$_{1-x}$P$_x$)$_2$(*10*) systems. Triangles indicate the compositions studied in this work. $T^*$ indicates the temperature at which the system starts to show $C_2$-symmetric physical properties.(*5,24*). PM, paramagnetic; SC, superconducting; Tetra., tetragonal; Ortho., orthorhombic. (**C1**) Schematics of four-fold symmetric band structure around Γ of IBSCs (*12*). For the photoemission with low photon energy (4.66 eV), the observable energy-momentum region is limited around the Fermi level ($E_F < E < 150$ meV) and Γ (from –0.2 Å$^{-1}$ < $k$ < 0.2 Å$^{-1}$) as indicated by the gray area. (**C2**) Schematics of the partial density of states for *xz* and *yz* orbtials around Γ in the tetragonal ($C_4$) and nematic ordered state ($C_2$). (**D** and **E**) LD images obtained by the laser linearly polarized along the orthorhombic axes for FeSe at 11 K and BaFe$_2$(As$_{0.87}$P$_{0.13}$)$_2$ at 65 K, respectively. Thick dashed lines represents the boundary at which the stripe patterns are rotated by 90 degrees. (**F** and **G**) LD images and integrated LD signals of rectangles 1 and 2 in (E),

respectively. We analyzed the integrated LD signals in the gray regions as indicated in Fig. 2, D and E.

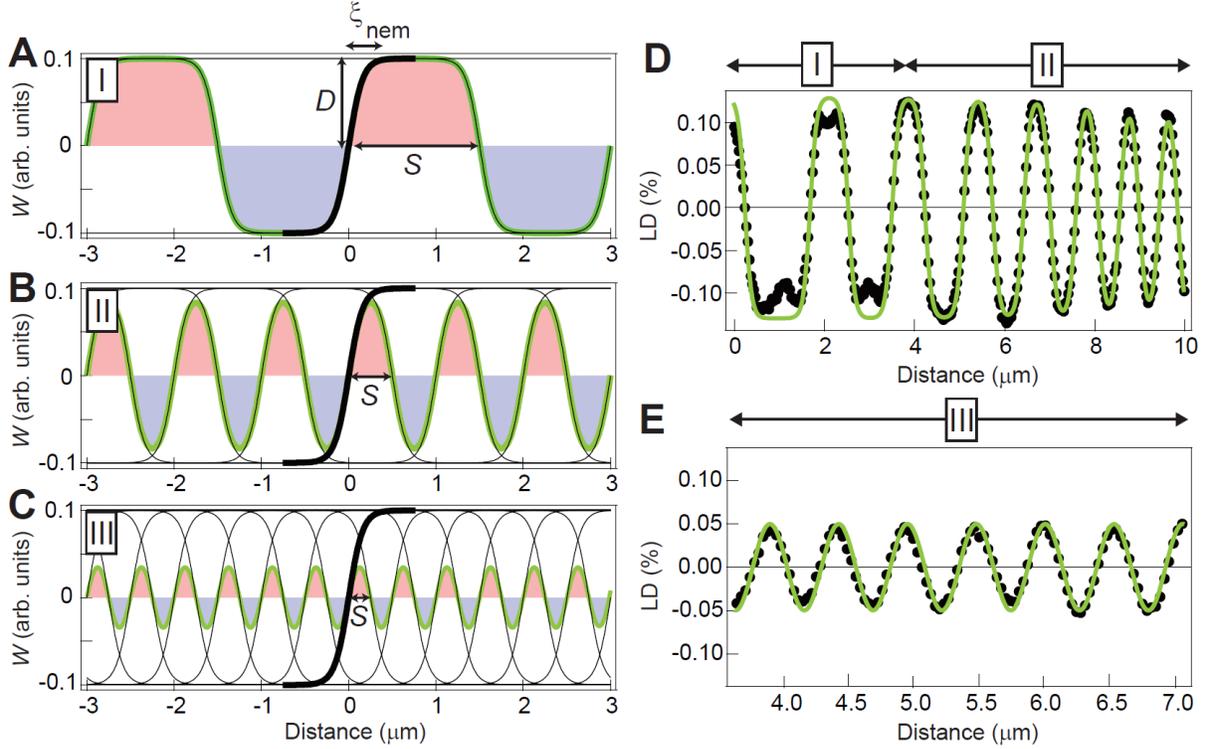

**Fig. 2. Fitting analysis of the LD signals in BaFe$_2$(As$_{0.87}$P$_{0.13}$)$_2$.** (**A** to **C**) Simulations of the LD signals (green curves) based on a model that assumes a train of nematic domain walls $W(x)=D\tanh[x/2\xi_{nem}]$ (black curves) with domain-wall interval $S$ = 1500, 500 and 250 nm, respectively. $D$ and $\xi_{nem}$ are set to 0.1 and 500 nm, respectively. (**D**) Fitting analysis on the LD signal in the gray region of Fig. 1F. The green curve represents the fitting function composed of the nematic domain walls with $D$ = 0.13, $\xi_{nem}$ = 450 nm, and $S(x)$ ranging from 400 to 1400 nm.

(**E**) The same as (**D**) but in the gray region of Fig. 1G. The parameters were set to $D = 0.13$, $\xi_{nem}$ = 450 nm, and $S = 260$ nm.

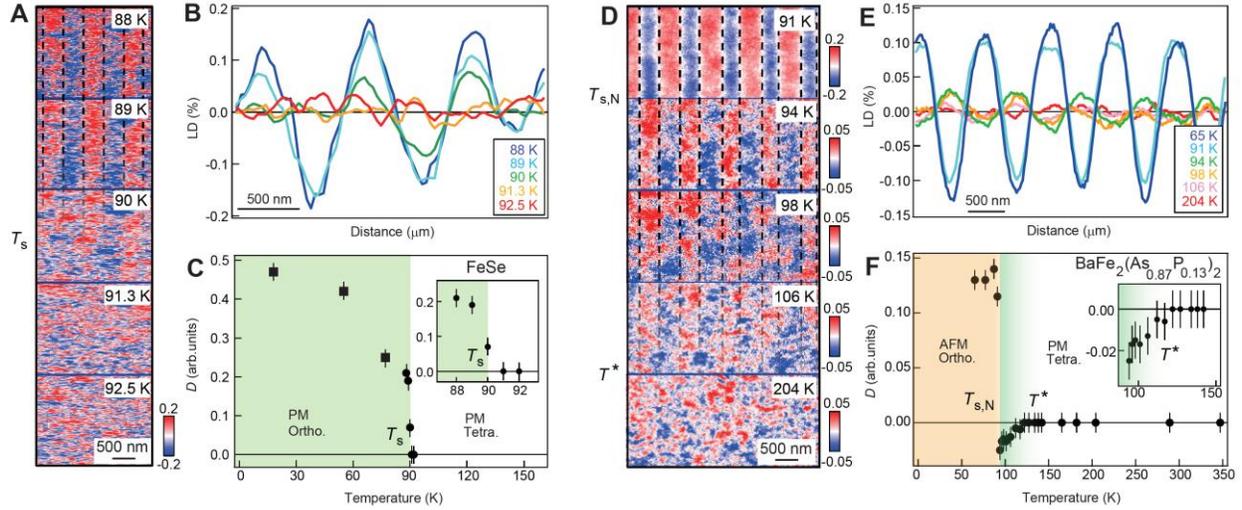

**Fig. 3. Temperature dependence of the LD signal.** (**A** and **B**) Temperature dependence of the LD image (A) and integrated LD signal across $T_s$ (= 90 K) (B) for FeSe. Dotted lines in (A) indicate the nodal positions of the nematicity wave. (**C**) Temperature dependence of the nematic-domain-

wall height (*D*) extracted from the fitting analysis. (**D** to **F**) The same as (A to C) but for BaFe$_2$(As$_{0.87}$P$_{0.13}$)$_2$ ($T_{s,N}$ = 93 K).

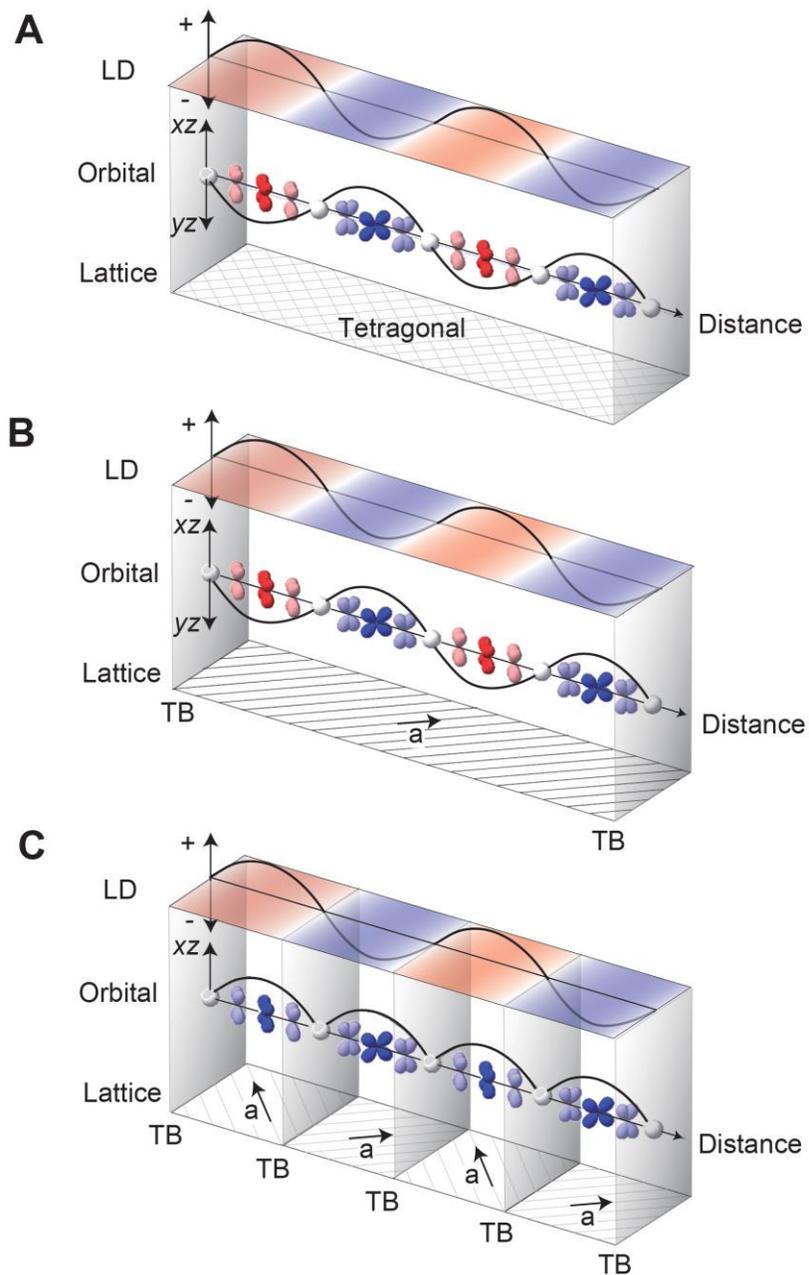

**Fig. 4. Real-space scenarios for the sinusoidal LD signals.** (**A**) Schematics for the sinusoidal LD signal at $T_s < T < T^*$. Blue and red orbitals represent the $xz$- and $yz$-dominated electronic states at $\Gamma$, respectively. (**B** and **C**) Same as (A) but at $T < T_s$ for the cases with and without correspondence between the locations of TBs and nodes of sinusoidal LD signal, respectively. The orthorhombic axes are rotated by 90 degrees at the TB. $x$, $y$, and $z$ are coordinates along the crystal axes of the orthorhombic setting $a$, $b$, and $c$, respectively.

# Supplementary Materials for
# Discovery of mesoscopic nematicity wave
# in iron-based superconductors

T. Shimojima[1,†,*], Y. Motoyui[2,†], T. Taniuchi[2,3], C. Bareille[2,3], S. Onari[4], H. Kontani[4], M. Nakajima[5], S. Kasahara[6,+], T. Shibauchi[7], Y. Matsuda[6], S. Shin[2,3,8,*]

[1]RIKEN Center for Emergent Matter Science (CEMS), Wako 351-0198, Japan
[2]Institute for Solid State Physics (ISSP), The University of Tokyo, Kashiwa, 277-8581, Japan
[3]Material Innovation Research Center (MIRC), The University of Tokyo, Kashiwa, Chiba 277-8561, Japan.
[4]Department of Physics, Nagoya University, Furo-cho, Nagoya 464-8602, Japan
[5]Department of Physics, Osaka University, Toyonaka, Osaka, 560-0043, Japan
[6]Department of Physics, Kyoto University, Kyoto 606-8502, Japan
[7]Department of Advanced Materials Science, The University of Tokyo, Kashiwa, 277-8561, Japan
[8]Office of University Professor, The University of Tokyo, Kashiwa, Chiba 277-8581, Japan

*Correspondences

†Equal contributions

+Present address: Research Institute for Interdisciplinary Science, Okayama University, 3-1-1 Tsushimanaka, Kita-ku, Okayama 700-8530, Japan.

**This PDF file includes:**
Materials and Methods
Supplementary Text
Figs. S1 to S9
References

## Materials and Methods:

### Laser-PEEM measurements

We used linear dichroism (LD) imaging in laser-PEEM at ISSP, University of Tokyo, to map the nematic order parameter at the crystal surface of under-doped AsP122 and FeSe. The light source is a continuous-wave laser of 4.66 eV which allows PEEM imaging with a high spatial resolution up to 2.6 nm(*21,36*). We chose the photon energy slightly higher than the work function of the IBSCs (~4.5 eV) in order to obtain the orbital polarization near the Fermi level around $\Gamma$ point(*21,36*). In this work, the spatial resolution was set to <90 nm in order to obtain higher signal-to-noise ratio (see SM, section 10). The light incidence was normal to the sample surface. The light polarizations were set parallel and perpendicular to the orthorhombic axes. Here, linearly polarized laser picks up the orbital that elongates along the same direction of light polarization. Single crystals were cleaved in situ at room temperature in an ultrahigh vacuum better than $2 \times 10^{-10}$ Torr. By considering the escape depth of the photoelectrons(*37*), the out-of-plane thickness of the LD pattern is estimated to be >100 nm.

### Sample preparations

High-quality single crystals of FeSe were grown by the vapor transport method. A mixture of Fe and Se powders was sealed in an evacuated $SiO_2$ ampoule together with KCl and $AlCl_3$ powders(*9*). The transition temperatures of the single crystals were estimated to be $T_s$ = 90 K and $T_c$ = 9 K from the electrical resistivity measurements. High-quality single crystals of $BaFe_2(As_{0.87}P_{0.13})_2$ ($T_{s,N}$ = 93 K) were grown by a self-flux method(*10*). Resistivity measurements suggest the high quality of the crystals with a residual resistivity ratio of ≤30.

## Supplementary sections:

### 1. Electronic structure around $\Gamma$ of IBSCs

The schematic band structure in Fig. 1C1 shows that the *xz/yz* orbital bands are degenerate at $\Gamma$. However, the detailed ARPES data(*13,15,38*) indicates the energy gap of ~20 meV between *xz/yz* bands at $\Gamma$ below and above $T_s$. In addition, the anisotropic band dispersions for *xz/yz* orbitals develop below $T_s$, indicating the emergence of the orbital polarization. Such temperature (*T*)-dependent band structure for FeSe was reproduced by the band calculations(*13*) including both *T*-independent spin-orbit interaction (SOI) (20 meV) and *T*-dependent orbital polarization (10 meV). Here, the band splitting due to SOI does not contribute to the LD signal since it cannot induce any inequivalence between the density of states for *xz* and *yz* orbitals. On the other hand, the orbital polarization is directly detected by the LD-PEEM. For simplicity, we thus omitted the energy gap due to SOI from the schematics in Fig. 1C1.

### 2. LD images in a large area for $BaFe_2(As_{0.87}P_{0.13})_2$.

Here we show seven LD images around the boundaries at which the LD stripes are rotated by 90° (thick white lines) in Fig. S1A. We found that the stripe patterns cover all of the field of views. Figure S1B shows the period and amplitude of the LD stripes as a function of the distance from the LD boundary (indicated by broken white arrow in Fig. S1A) obtained from the rectangle areas in the five fields of views (A-E) in Fig. S1A. We confirmed the tendency in a large area of the crystal surface that the LD signal shows lower amplitude where the wavelength is shorter.

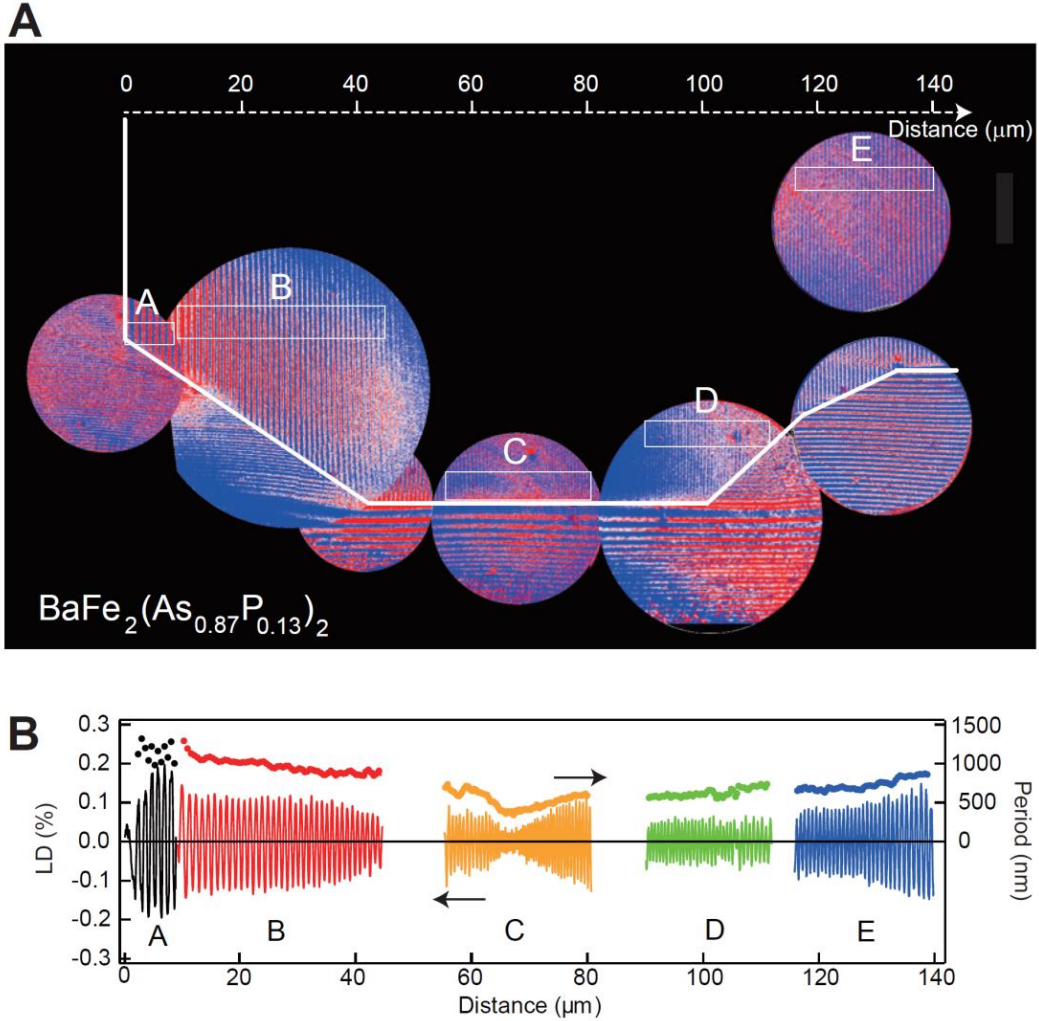

**Fig. S1: LD images in a large area for BaFe$_2$(As$_{0.87}$P$_{0.13}$)$_2$**
(**A**) Seven LD images obtained at 65 K from the same crystal surface of BaFe$_2$(As$_{0.87}$P$_{0.13}$)$_2$. Thick white lines represent the boundaries at which the LD stripes are rotated by 90°. (**B**) Spatial variations of the amplitude (solid curves) and period (filled circles) of the LD stripes obtained from the selected areas A-E [white rectangles in (A)]. The origin and direction of the bottom axis are indicated by broken white arrow in (A).

### 3. LD image and integrated signals for FeSe
In Fig. S2A, we show the LD image for FeSe obtained at 11 K. Thick dashed lines represent the boundaries at which the LD stripes are rotated by 90°. We find that the complex LD pattern is resemblance to that for BaFe$_2$(As$_{0.87}$P$_{0.13}$)$_2$ (Fig. 1E). Figure S2B shows the LD signal integrated in the rectangle in Fig. S2A. As shown in Fig. S2C, the LD signal in the gray area in Fig. S2B can be well reproduced by the train of the nematic domain walls written by the hyperbolic tangent functions with $D = 0.11$, $\xi_{nem} = 550$ nm and $S = 350$ nm (green curve), which is nearly equivalent to the sinusoidal wave. The period of the oscillating LD signal varies in the field of view in Fig. S2A.

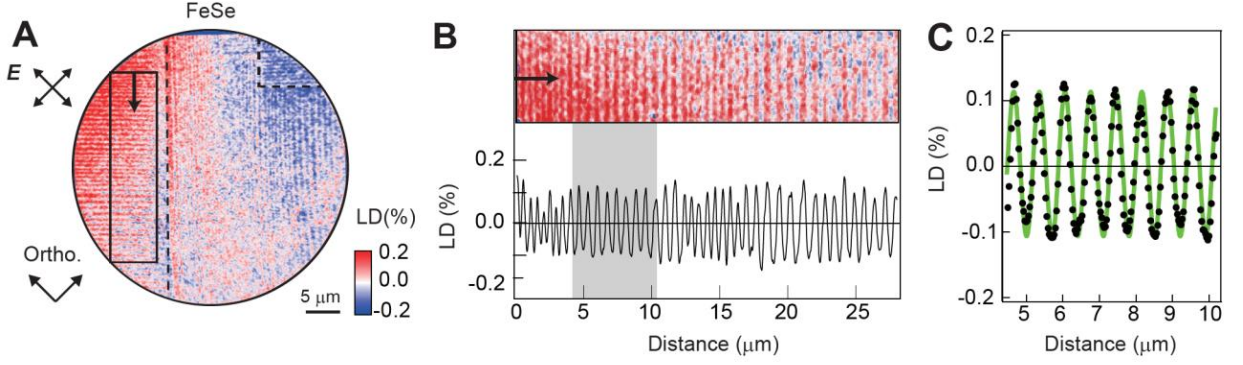

**Fig. S2 Fitting analysis on the LD data for FeSe** (**A**) LD image for FeSe obtained at 11 K. (**B**) LD image and integrated LD signal in the rectangle in (A). (**C**) Fitting result on the LD signal in a gray area in (B) by using the train of the nematic domain walls with $D = 0.11$, $\xi_{nem} = 550$ nm and $S = 350$ nm (green curve).

## 4. Ginzburg-Landau theory for nematic order

Here, we discuss the $r$-dependence of the nematic order parameter $\phi(r)$ around the nematic domain wall. The Ginzburg-Landau free-energy is given as

$$F = \int \left( a\phi(r)^2 + \frac{b}{2}\phi(r)^4 + c|\nabla\phi(r)|^2 \right) dr$$

where $a \propto T_c - T$ ($T_c$ is the critical $T$), and both $b$ and $c$ are positive. Here, we assume that single domain parallel to $y$-axis is located at $x = 0$, and $\phi$ depends only on $x$. When $\phi(x) = +(-)\phi_0$ for $x = +(-)\infty$ for $T < T_c$, the solution of the Euler equation is given as

$$\phi(x) = \phi_0 \tanh\left(\frac{x}{2r_c}\right)$$

where $\phi_0 = \sqrt{|a|/b}$, and $r_c = \sqrt{2c/|a|}$. Thus, the width of domain wall $r_c$ becomes long when $c$ is large.

## 5. Spatial distribution of S value

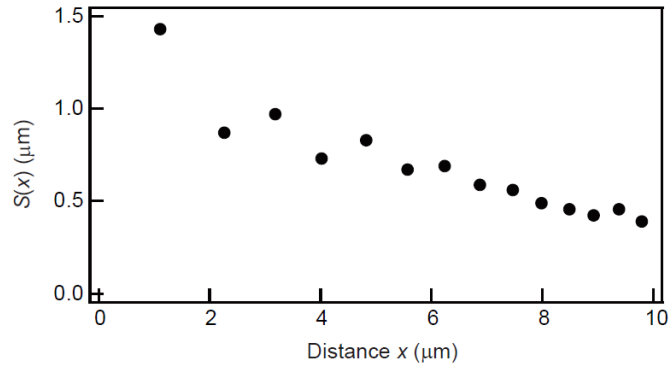

**Fig. S3.** Spatial distribution of $S(x)$ in Fig. 2D.

Figure S3 shows the $S(x)$ values for the nematic domain walls obtained from the fitting analysis in Fig. 2D. The bottom axis corresponds to that of Fig. 2D whose origin is set to the nematic domain boundary (Fig. 1, E and F). $S(x)$ was plotted at $x$ corresponding to the halfway point between two nematic domain walls which define $S(x)$. We found that $S(x)$ gradually increases as the nematic domain wall approaches the nematic domain boundary.

## 6. Fitting analyses on the temperature-dependent LD signals

We analyzed the LD signals in order to obtain the nematic-domain-wall height ($D$) in a wide $T$ range. Figures S4, A to H, show the $T$ dependence of the LD signals from 18 K to 92.5 K for FeSe. We used the dataset 1 for Fig. S4, A to C, (18 K - 77 K) and dataset 2 for Fig. S4, D to H, (88 K - 92.5 K) which were obtained from different field of views on the same cleaved crystal surface. We fitted the slope of the LD signals by using the formula $W(x)=D\tanh[x/2\xi_{nem}]$ (red curves). By setting $\xi_{nem} = 550$ nm for all temperatures, we obtained $D$ values as indicated in the bottom of each panel. $D$ values at 91.3 K and 92.5 K were set to zero within the error bar since the sinusoidal waveform was not clearly observed. We plotted the $T$ dependence of $D$ for FeSe in Fig. 3C.

In a similar way, by setting $\xi_{nem} = 450$ nm for all temperatures, we obtained $D$ values from 65 K to 134 K for $BaFe_2(As_{0.87}P_{0.13})_2$ as shown in Fig. S5, A to N. Note that the sign of $D$ changes between 91 K and 94 K. For the temperatures from 122 K to 350 K (not shown), $D$ values were set to zero within the error bar. Finally, we obtained the $T$ dependence of $D$ for $BaFe_2(As_{0.87}P_{0.13})_2$ as shown in Fig. 3F.

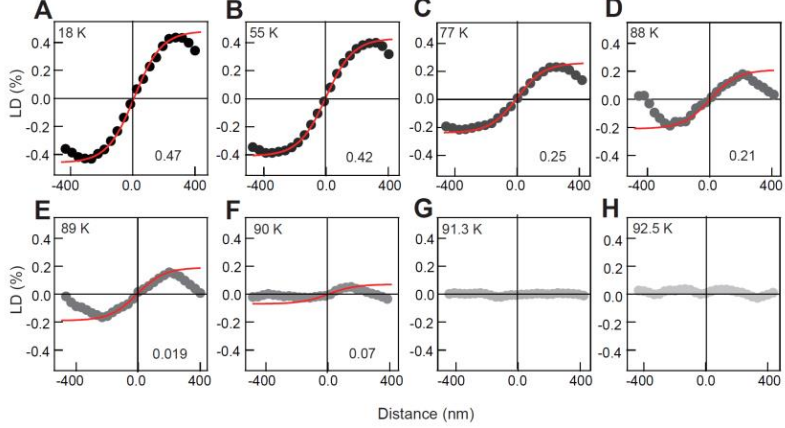

**Fig. S4. Fitting analyses on the LD signals in a wide temperature range for FeSe.** (**A** to **H**) The LD signals from 18 K to 92.5 K with fitting curves written by $W(x)=D\tanh[x/2\xi_{nem}]$ (red curves), respectively. Obtained $D$ value was indicated at the bottom of each panel.

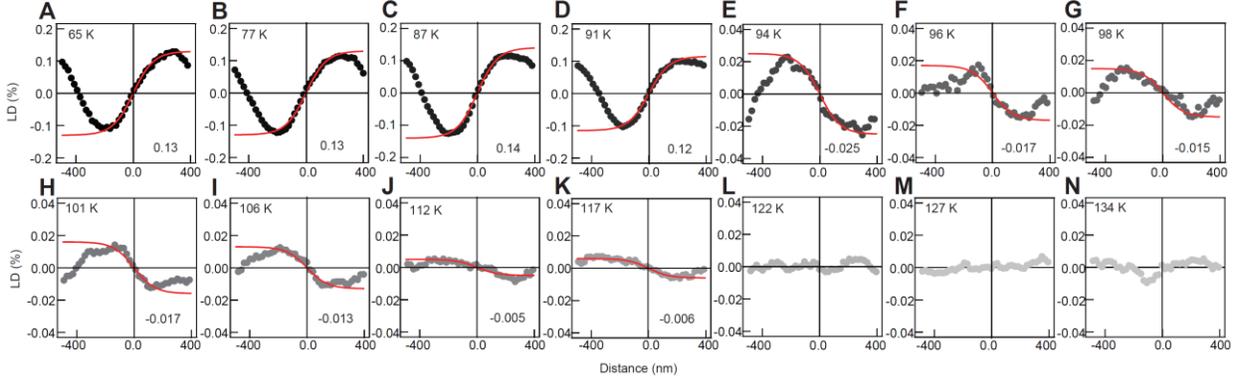

**Fig. S5. Fitting analyses on the LD signals in a wide temperature range for $BaFe_2(As_{0.87}P_{0.13})_2$.** (**A** to **N**) The LD signals from 65 K to 134 K with fitting curves written by $W(x)=D\tanh[x/2\xi_{nem}]$ (red curves), respectively. Obtained $D$ value was indicated at the bottom of each panel.

## 7. Temperature-dependent Laser-ARPES at Γ for detwinned FeSe

For investigating the $T$ dependence of the orbital polarization around Γ, we performed laser-ARPES(*39*) on detwinned single crystals of FeSe at ISSP, University of Tokyo. Figures S6, A and B, show the band dispersions along $k_y$ obtained at 120 K and 30 K, respectively. In the nematic ordered state ($T < 90$ K), the Fermi momenta of the hole band around Γ expand along $k_y$ and contract along $k_x$, thus forming an elliptical Fermi surface (inset of Fig. S6B) (*13*). As $T$ decreases below 90 K, we observed the shift of the hole band to the right-hand side as indicated by the white arrows in Fig. S6, A and B. To see the $T$ dependence more clearly, we show the momentum distribution curves (MDCs) at the binding energy 10 meV (dotted line in Fig. S6, A and B) in Fig. S6C. The peak positions (black arrows) correspond to the location of the hole band which clearly indicates the shift below $T_s = 90$ K. In Fig. S6D, we plotted the shift of the hole bands (i.e. amplitude of the orbital polarization) as a function of $T$. It clearly shows a rapid increase below $T_s$ which is resemblance to the $T$ dependence of $D$ in Fig. 3C.

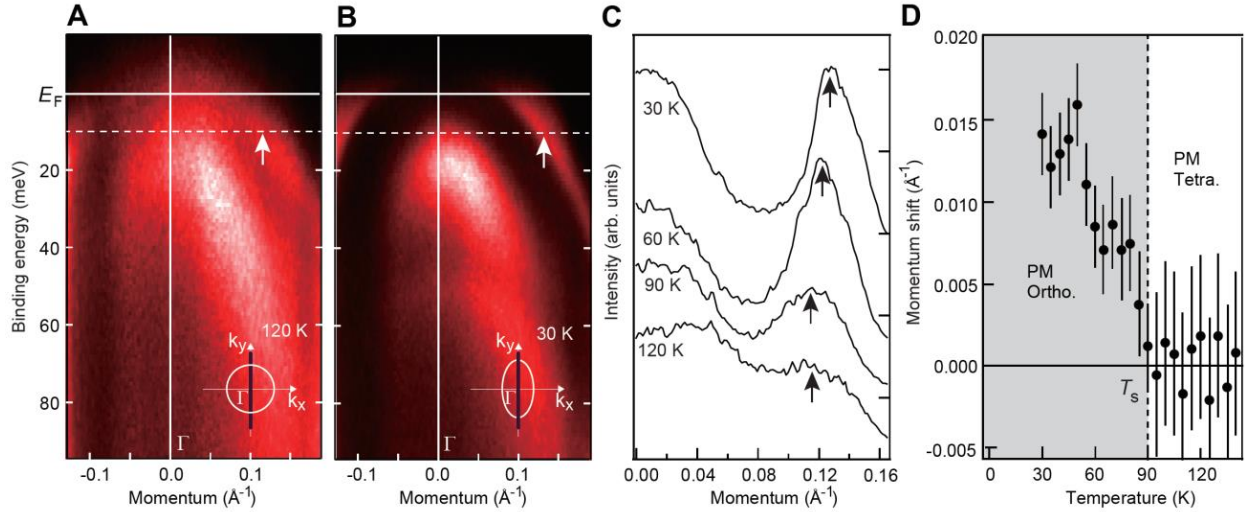

**Fig. S6: Temperature-dependent laser-ARPES on detwinned FeSe**
(**A** and **B**) Band dispersions along $k_y$ for detwinned single crystals of FeSe obtained by laser-ARPES with *p*-polarized laser at 120 K and 30 K, respectively. Insets show the schematic Fermi surfaces (white circle and ellipse) and momentum cuts for laser-ARPES (black lines). (**C**). Momentum distribution curves at the binding energy 10 meV [dotted lines in (A) and (B)] for 30 K, 60 K, 90 K and 120 K. (**D**) Detailed *T* dependence of the MDC peak positions.

## 8. Observation of the nematicity wave around the crystal defects

In order to understand the relation between the nematicity wave formation and the local strain, we conducted LD-PEEM experiments near the crystal defects (deep cracks) which were possibly generated during the sample growth. Figure S7A shows the PEEM image for FeSe at 60 K including the cracks, as indicated by the black lines in Fig. S7C. In the LD-PEEM image of the same field of view (Fig. S7B), we found that the nematic domain boundary at which the LD stripes are rotated by 90° (dotted black lines in Fig. S7C) does not correspond to the location of the cracks. The LD stripes in the left-hand side of Fig. S7B are aligned parallel to the crack. On the other hand, in a different field of view (Fig. S7, D to F), the LD stripe is nearly perpendicular to the crack. These observations suggest no correspondence between the directions of the cracks and LD stripes. By assuming that these linear cracks induce uniaxial strain field on the sample surface, we can consider that the nematicity wave formation is not originated from the local strain. While the information of the spatial distribution of the strain field is still limited on a nanometer scale, these LD-PEEM data suggest that the nematicity wave is intrinsically electronic phenomenon in the iron-based superconductors.

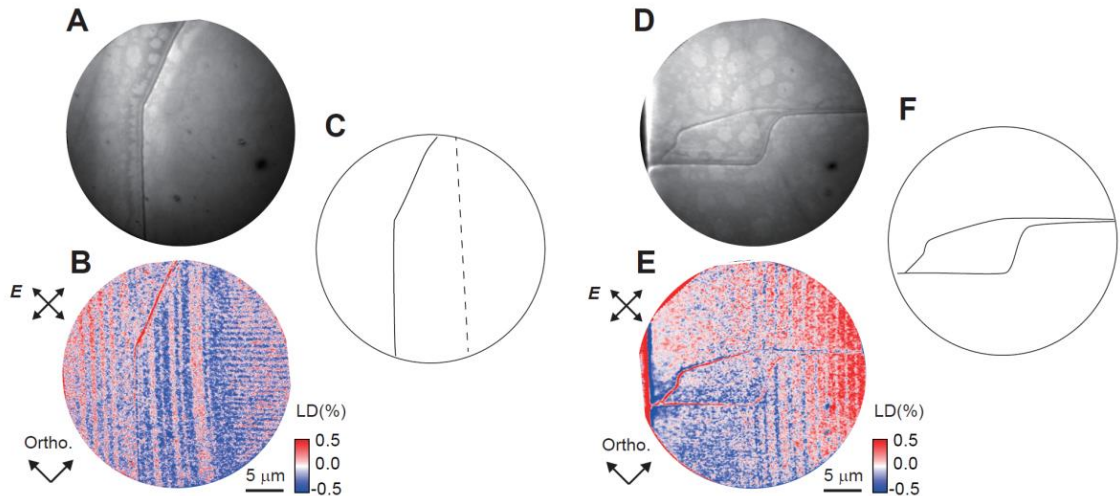

**Fig. S7** (**A** and **B**) PEEM image and LD-PEEM image for FeSe at 60 K, respectively. (**C**) Schematics of the crystal defect (black lines) and the nematic domain boundary at which the LD stripes are rotated by 90° (dotted black line). (**D** to **F**) The same as (A to C) but in a different field of view.

## 9. Temperature-cycle measurements of LD-PEEM

We conducted temperature ($T$)-cycle LD-PEEM measurements in a fixed field of view. Figures S8, A and B show the LD-PEEM images for FeSe obtained at 70 K and 80 K, respectively. Note that the latter was obtained after the $T$ cycle across $T_s$ = 90 K (70 K → 100 K → 80 K). Black arrows indicate the locations of the impurities on the sample surface which ensure the common field of views in Fig. S8, A and B. We found a drastic change in the LD pattern after the $T$ cycle. As demonstrated for the strain-induced metal-insulator-phase modulation in ref. *40*, we can expect that the LD stripes will be repeatably aligned in a fixed direction if the uniaxial tensile or compressive stress originating from the mounting/cooling processes is applied to the sample. However, this is not the case with the present LD-PEEM results. We thus consider that the observation of the unrepeatable formation of the LD patterns suggests a negligible influence from such extrinsic strains.

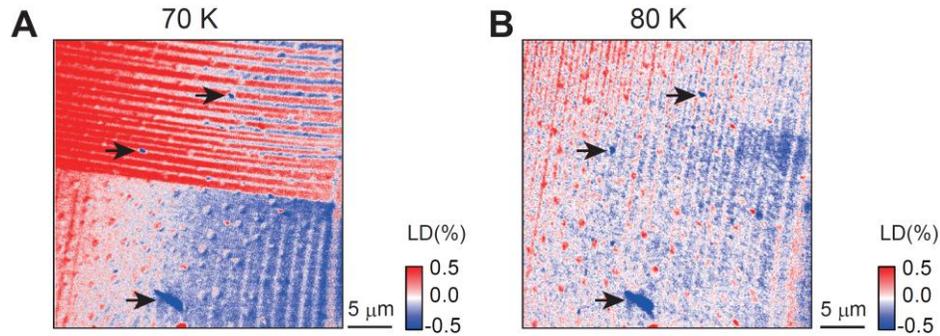

**Fig. S8** (**A**) LD-PEEM images for FeSe obtained at 70 K. (**B**) The same as (A) but at 80 K obtained after temperature cycling across $T_s$ = 90 K. Black arrows indicate the locations of the impurities on the sample suggesting the common field of views for (A and B).

## 10. Spatial resolution of Laser-PEEM

The spatial resolution was estimated from the fitting analysis on the crystal defects in the PEEM image. Since the line profile of the defect is assumed to follow the step function, we can estimate the upper limit of the spatial resolution. As shown in Fig. S9A, we reproduced the dip structure in the line profile of the defect (inset of Fig. S9A) by using the step functions convoluted by the gaussian with the full width at half maximum (FWHM) of 90 nm. We then obtained a spatial resolution of <90 nm.

Here we show that the sinusoidal waveform of the LD signal is not caused by the resolution effect on the rectangle wave. In Fig. S9B, we show the LD signal with the oscillation period of 530 nm (black markers) which is well reproduced by the sinusoidal wave (green curve). The red curve represents a rectangle wave with a period of 530 nm (black lines) broadened by the spatial resolution of 90 nm. The broadened rectangle wave (red curve) is clearly distinguished from the LD signal suggesting that the sinusoidal waveform is an intrinsic feature of the LD signal. This result holds for the sinusoidal LD signals with the wavelength longer than ~500 nm.

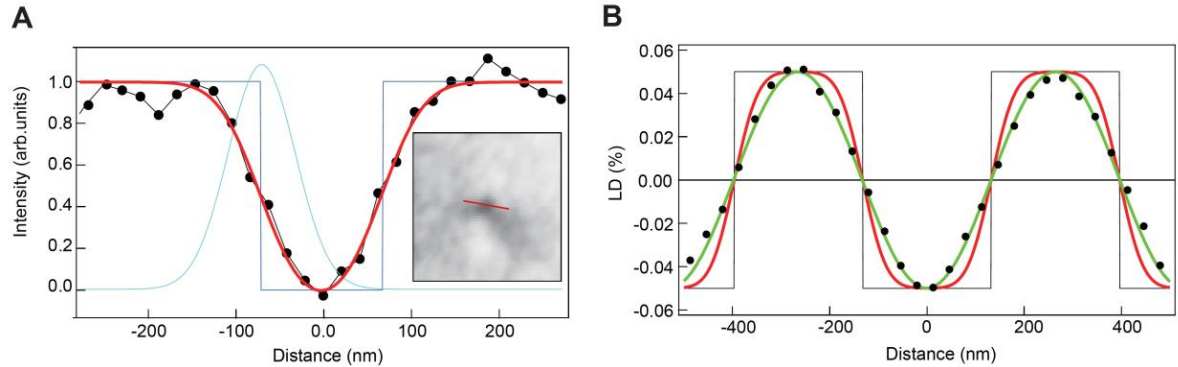

**Fig. S9 Estimation of the spatial resolution for laser-PEEM**. (**A**) Black markers represent the profile of the defect along the red line in the inset. Red curve represents the fitting function composed of the step functions (blue line) convoluted by the gaussian with the FWHM of 90 nm (light blue curve). (**B**) Fitting results on the LD signal (black markers) in Fig. 1G. Green and red curves represent a sinusoidal wave with a wavelength of 530 nm, and a rectangle wave with the same period convoluted by the gaussian with the FWHM of 90 nm, respectively.